%% file: FNunez_Perez-Acle_etal2012.tex
\documentclass[10pt]{article}

\usepackage{amsmath}
\usepackage{amssymb}

\usepackage{graphicx}

\usepackage{cite}

\usepackage{color}

\usepackage[labelfont=bf,labelsep=period,justification=raggedright]{caption}

\makeatletter
\renewcommand{\@biblabel}[1]{\quad#1.}
\makeatother

\date{}

\begin{document}

\begin{flushleft}
{\Large \textbf{A Rule-based Model of a Hypothetical Zombie Outbreak: Insights on the role of emotional factors during behavioral adaptation of an artificial population} }
\\
Felipe Nu\~nez$^{1}$, Cesar Ravello$^{1}$, Hector
Urbina$^{1}$, Tomas Perez-Acle$^{1,2\ast}$
\\
\bf{1} Computational Biology Lab (DLab), Fundaci\'on Ciencia \& Vida\\
Avenida Za\~nartu 1482, \~Nu\~noa, Santiago, Chile\\

\bf{2} Centro Interdisciplinario de Neurociencias de Valpara\'iso, Universidad de Valpar\'iso\\
Valpara\'iso, Chile\\

$\ast$ Corresponding author: tomas@dlab.cl
\end{flushleft}

\section*{Abstract}
\input{abstract}

\section*{Introduction}
\input{introduction}

\section*{Models}
\input{methods}

\subsection*{Implementation}
\input{implementation}

\subsection*{Parameters}
\input{parameters}

\subsection*{Kappa file}
\input{prekappa}

\subsection*{Simulations}
\input{simulations}

\section*{Results}
\input{results}

\section*{Discussion}
\input{discussion-model}

\input{discussion-kappa}

\input{discussion-results}

\input{conclusion-perspective}

\section*{Acknowledgments}
To Ricardo Honorato-Zimmer and Sebastian Gutierrez for extensive discussions on these matters. We acknowledge partial financing support from Proyecto de Financiamiento Basal Fundacion Ciencia para la Vida (PFB03), and Centro Interdisciplinario de Neurociencias de Valparaiso (ICM-ECONOMIA P09-022-F). We acknowledge supercomputing time from the National Laboratory for High Performance Computing NLHPC (ECM-02). Powered@NLHPC.
\bibliography{FNunez_Perez-Acle_etal2012_biblio}

\input{figures}

\input{tables}

\end{document}

%% file: abstract.tex
Models of infectious diseases have been developed since the first half of the twentieth century. There are different approaches to model an infectious outbreak, especially in terms of how individuals and their interactions are defined and treated. Most models haven't considered the role that emotional factors of the individual may play on the population's behavioral adaptation during the spread of a pandemic disease. Considering that local interactions among individuals generate patterns that -at a large scale- govern the action of masses, we have studied the behavioral adaptation of a population induced by the spread of an infectious disease. Therefore, we have developed a rule-based model of a hypothetical zombie outbreak, written in \emph{Kappa} language, and simulated using Guillespie's stochastic approach. Our study addresses the specificity and heterogeneity of the system at the individual level, a highly desirable characteristic, mostly overlooked in classic epidemic models. Together with the basic elements of a typical epidemiological model, our model includes an individual representation of the disease progression and the traveling of agents among cities being affected. It also introduces an approximation to measure the effect of panic in the population as a function of the individual situational awareness. In addition, the effect of two possible countermeasures to overcome the zombie threat is considered: the availability of medical treatment and the deployment of special armed forces. However, due to the special characteristics of this hypothetical infectious disease, even using exaggerated numbers of countermeasures, only a small percentage of the population can be saved at the end of the simulations. As expected from a rule-based model approach, the global dynamics of our model resulted primarily governed by the mechanistic description of local interactions occurring at the individual level. As a whole, people's situational awareness resulted essential to modulate the inner dynamics of the system.

%% file: introduction.tex
Zombies are fictitious entities described in tales throughout history as human beings that, through various methods, have passed from a cataleptic state to a pseudo-life, lacking self control \cite{zombihaiti, survival}. The etymological origin of the Zombie word can be traced to the Voodoo cult, in which, according to the controversial work of Davis\cite{davis1983,booth1988}, a cataleptic person -induced by a toxin or venom- could be raised from the grave by a wizard to be turned into his slave. However, the actual and more popular concept is from films of George A. Romero, among others, in which a zombie is a human affected by a highly contagious disease -typically a virus- that turns him into a mindless and wandering being with an insatiable hunger for human flesh \cite{romero1, romero2}. According to pop culture, zombies represent entities that generate social chaos, similar to large scale outbreaks of infectious diseases, leading to a state of catastrophe, affecting people physical and emotionally \cite{survival, romero2, walkingdead}. Thus, the imaginary zombie scenario dictates that societies suffering an outbreak are inevitably driven to a loss of control and misrule. In this setting, decisions are taken by considering only local and immediate information elements regarding the situation to be surpassed, mainly focused on surviving\cite{romero2, walkingdead}. At the governmental rank, decisions are taken by considering the general situation and military forces are often deployed to combat the zombie horde. Additional measures such as quarantine \cite{legend} and the use of weapons of mass destruction \cite{resident}, are typical approaches exploited in pop culture to try to solve this menacing problem. In the few situations where treatment is available\cite{legend}, logistic problems usually disrupt the effective delivery of treatment to infected people. On the other hand, the topology of connectivity between cities is not always considered when imposing quarantine countermeasures, nor can they be sufficient to retain infected people to travel from one city to the other. Moreover, it is always possible, but less likely, to suffer from a multiple zombie outbreak, making the quarantine countermeasures even harder to be sustained. In spite of the desperate efforts to combat the zombie horde, these usually win. A 100\% effective transmission upon bite, a short incubation period, the uncontrollable desire of zombies to infect, no cure or remission, and people's ignorance to deal with such an unexpected situation, together with a profound emotional outcome, configures a situation where the survival of the human race is a complex task.

Many infectious diseases have models that describe how they spread on the population, and the case of a zombie outbreak is not an exception.  Munz \emph{et al.} \cite{zombieode} presented a simple model based on ordinary differential equations (ODE) with perfect mixing that follows the mass action kinetics, closely related to classic epidemiological descriptions as the SIR (susceptible - infected - recovered) model \cite{SIR}. Later on, Crossley \emph{et al.} \cite{simzombie} translated Munz model, almost directly without further improvements, to an agent-based model (ABM). An ABM that considers the heterogeneity of agents and space is an important step forward to include relevant features commonly overlooked in simpler models.  In spite that ABM simulation offers a set of new possibilities to study a zombie outbreak, it is arguable -philosophically speaking- that zombies, or even people, actually have an intrinsic purpose \cite{pross,monod}. To study the influence that individual behavioral adaptation may impose on the internal dynamics of populations being affected by the spread of an infectious disease, we have implemented a rule-based model of a hypothetical zombie outbreak. Our model proposes that individuals affected by catastrophic events, specially zombies, have no intrinsic purpose and that at a certain scale, the behavior of people -agents in the simulation- follow patterns beyond their selves. Considering that agents in a rule-based model should express the inner heterogeneity of the system, we decided to describe the dynamics of a zombie infection by using a highly comprehensive tool such as the \emph{Kappa} language \cite{danos2004}. \emph{Kappa} provides a formalism to define agents that interact with each other, according to general rules, through interfaces composed by a set of sites. Each site always have internal states and can have a binding state, which maps to a site belonging to another agent. The internal state of a site is a label that indicates a certain local state of the agent. In the case of our zombie model, the binding states are used to describe encounters between agents during the simulation and also to represent the progression of different phenomena, such as infection growth or distance travelled by agents. On the other hand, internal states are used to say if an agent is a susceptible, infected, or zombie person, to represent his level of panic, the city where the agent resides, etc. In order to deal with the combinatorial complexity arising from the heterogeneity of agents, \emph{Kappa} rules are defined as patterns representing only the interesting parts of the interactions between agents and/or complexes. Simulations are performed using the Gillespie's Stochastic Simulation Algorithm (SSA) \cite{gillespie}. According to this Gillespie's based approach, systems to be simulated are defined as a mixture of agents moving randomly in a given space, generating collisions between them that can lead to an interaction, defined as a \emph{Kappa} rule \cite{danos2007}. Using this rule-based model, we have explored the case of a hypothetical zombie outbreak, by incorporating the basic elements of a typical epidemiological model. We have included a representation of the disease progression for every infected individual, and a novel approximation to measure the panic effect in the behavioral adaptation of the population, as a function of the individual situational awareness. Furthermore, we have included the effect of panic in the movement of people among infected cities, and the effect of medical treatment and the arrival of special trained forces to kill zombies. Altogether, this complex scenario allowed us to study the role that emotional factors on the individual may play on the population's behavioral adaptation.

%% file: methods.tex
Following the zombie pop culture and the classical SIR model of infectious diseases \cite{SIR}, we can define a zombie \emph{(Z)} as the vector of an infectious disease. \emph{Z} is a person that is capable of transmitting the disease to susceptible individuals \emph{(S)} but, on the contrary of asymptomatic vectors, they also expresses the disease symptoms. For the infection to occur, an \emph{S} must make direct contact with body fluids of \emph{Z} \cite{survival}, so, there must be an encounter that results in the defeat of \emph{S}, typically manifested as \emph{S} being bitten by \emph{Z} (Figure \ref{sidzet-edo}). The actual disease is usually defined by three stages \cite{romero1,romero2,survival}. The first stage is infection (\emph{I}) that represents the incubation period and is characterized by an asymptomatic phase followed by a rapid decay of health. The second phase is characterized by a death-like state (\emph{D}), where the infected individual has no apparent vital signs but is suffering deep physiological and probably genetic changes induced by the pathogen. These set of changes forces to that individual to rise as \emph{Z}, the third and final phase of the disease. The transformation of an asymptomatic \emph{I} to \emph{Z} could be accelerated by another zombie attack: if \emph{I} is bitten by \emph{Z}, it immediately becomes \emph{D}.

On the other hand, an encounter with \emph{Z} could end in a victory for \emph{S}, or \emph{I}, by somehow destroying \emph{Z}, which turns it into a Removed (\emph{R}), \emph{i.e.} a person definitively dead in the sense that \emph{R} cannot rise as \emph{Z} again. The \emph{R} state is also reached by persons that died from circumstances that do not directly involve the effects of the disease, such as accidents and other natural causes. Another source of \emph{R} are \emph{D} individuals that have been prevented from becoming \emph{Z}, either by the action of \emph{S} or \emph{I}, or by being completely eaten by \emph{Z}. Finally, \emph{Z} individuals become \emph{R} after a long period of starvation.

To try to overcome the outbreak, we have included some individuals specialized in the removal of \emph{Z}, called Exterminators (\emph{E}), which are deployed at different times after the outbreak. Just like any other person, \emph{E} individuals can get the disease, but they turn into a infected exterminator (\emph{Ei}) instead of just \emph{I}, retaining their abilities to kill \emph{Z} more efficiently. Later on, just like \emph{I}, \emph{Ei} turn into \emph{D} and follow the aforementioned disease phases. We have also included treatment units \emph{T} that may, upon consumption, prevent the transition from \emph{I} and \emph{Ei} to \emph{D}, by turning them back into \emph{S} and \emph{E}, respectively.

In trying to address the effect of the psychological factor, namely panic, we defined that a person has three mental states. A higher level represents a person moving faster influenced by panic, so this person has a higher chance of an encounter, either with another person (\emph{S}, \emph{I}, \emph{E} or \emph{Ei}) or with a \emph{Z}, also having fewer chances of defeating \emph{Z} or obtaining $T$, due to diminished decision-making capabilities. A person can change its own mental state according to the situational awareness, for example, through an encounter with \emph{Z} and \emph{E}, or with another person. In an encounter with a \emph{Z}, the person will increase the level of panic if he/she becomes infected, or decrease it if he/she destroys \emph{Z}. In encounters with persons with a higher level of panic, the panic will increase. Someone with a high level of panic can spread the panic, but a person with a low level of panic cannot calm down others. If a person in the lowest mental state (\emph{p1}) encounters an \emph{E}, this person will increase his/her mental state to level \emph{p2}, but, if his/her mental state previous to the encounter is at level \emph{p3}, his/her mental state could decrease to \emph{p2}. The assumption behind this scheme is that an unaware person that suddenly encounters military personnel, would become worried, while conversely, a person in a state of panic would certainly become calm by feeling more protected. Also, an \emph{I} may increase his/her level of panic as the symptoms become apparent, while an \emph{S} may calm down spontaneously, given enough time without encountering with \emph{Z} or other persons in panic. We also included in our model an additional level, the zero level, that represents a fixed cold mind state that is achieved through previous successful experiences defeating \emph{Z}. In this state, a person has more chances to survive an encounter with a \emph{Z}, producing at the same time less encounters. People in level zero may calm down people with higher levels of panic. \emph{E} agents are supposed to be trained to control their fear and panic, so their mental state does not change in the model throughout the simulation.

Finally, looking for the description of large-scale problems, the model is extended into various compartments that represent different cities where people are able to travel (Figure \ref{cities-top}). On the contrary of \emph{S} and \emph{I}, \emph{Z} and \emph{E} cannot travel between cities. As \emph{E} individuals are specifically assigned to each city, they stay where they were deployed during the simulation. Despite \emph{Z} can move freely inside cities, they are not capable to interact with transporting agents (see below) during the simulation.

%% file: implementation.tex
Given the complexity of the described model, the implementation in \emph{Kappa} is not straightforward. We have defined an agent \emph{Person}, used to represent \emph{S,I,D,Z,E} and \emph{Ei} individuals. To distinguish them, the \emph{Person} agent has a site named $c$ (\emph{class}) whose internal state can be either one of the values $s$,$i$,$d$,$z$,$e$ or $ei$, as seen in Figure \ref{kappa-rule} (Panel A). Each \emph{Person} also has a site $i$ to interact with the $i$ site of other persons, a site $p$ to define its level of panic, a site $l$ to define its location and a site $v$ where to bind the virus after infection. All the interactions between agents involve the formation of complexes, thus changes in state of sites can only occur after the complex is formed. The separation of complexes occur spontaneously, independent of wether a reaction was executed or not. Moreover, there could be instances of two or more successive changes in the same complex, before separation takes place.

The infection process involves various changes in the \emph{S-Z} complex, as seen in Figure \ref{kappa-rule} (Panels B and C). Upon infection, the $c$ site's internal state of the \emph{Person} changed from $s$ to $i$, a \emph{Virus} agent is bound to its $v$ site and its panic rises one level (from $p_{1}$ to $p_{2}$ or from $p_{2}$ to $p_{3}$).
As a consequence of the preceding changes, the complex is separated so the result is a \emph{Z} , free to interact with another \emph{Person}, and an \emph{I} that will begin to develop the disease. The disease progression is represented as the growth of a chain of \emph{Virus}. When the growth of this chain reaches a certain length $n$, the infected \emph{Person} dies. It is important to note that these viruses are not meant to represent the progression of actual viruses in a host organism. On the contrary, this implementation is to avoid the exponential behavior of the interactions due to the SSA and to model the incubation period of an infectious disease, a process that differ largely from the mass-action regime. If the rate of replication of the virus $\delta$ is constant, we can consider that the infected goes through many phases according to the length of its chain of viruses, so the expression of the disease is given by:
\begin{equation}
  \begin{array}{rcll}
    I_1'(t)&=&-\delta I_1(t)&\\
    I_i'(t)&=& \delta I_{i-1}(t)-\delta I_i(t)&1<i<n\\
    I_n'(t)&=& \delta I_{n-1}(t)&\\
    I_1(0)&=&C_0&\\
    I_i(0)&=&0&1<i\leq n
  \end{array}
\end{equation}
Solving that system we can obtain an expression for $I_n(t)$:
\begin{equation}
I_n(t)=C_0\frac{\left(\delta t\right)^{n-1} e^{-\delta t}}{(n-1)!}
\label{eq-in}
\end{equation}
Then, the amount of infected persons in every time $t$ is determined by:
\begin{equation}
I(t)=C_0\sum_{k=0}^{n-2}\frac{\left(\delta t\right)^k e^{-\delta t}}{k!}
\label{eq-i}
\end{equation}

So if we want to model an incubation period of about $a$ to $b$ hours with an 80\% of the infected expressing the disease in that timespan, then we solve equation \ref{eq-i} to find the $\delta$ and $n$ to satisfy:
\begin{equation}
  \begin{array}{rcl}
    I(a)&=&0.9C_0\\
    I(b)&=&0.1C_0
  \end{array}
\label{eq-det}
\end{equation}

A similar approach was followed to implement the duration of travel among cities. We defined a \emph{Carrier} that binds to a \emph{Person} and to a \emph{Kilometer} that replicates forming a chain. The journey is completed when the chain of replicating \emph{Kilometer} reaches a certain length. In addition to the interaction sites with \emph{Person} and \emph{Kilometer}, \emph{Carrier} has also a site to define its city of origin and another one to define its destiny. Therefore, the length and replication rate of \emph{Kilometer} accounts for the duration of the journey. On the other hand, the number of \emph{Carrier} agents of each origin/destiny pair accounts for the capacity of each road, proportional to the number of lanes in each direction.

The effect of treatment also involves replicating chains. When a \emph{T} is consumed by an \emph{I} or an \emph{Ei}, an \emph{Antibody} is bound to the first \emph{Virus} of the chain. The \emph{Antibody} chain grows in parallel to the \emph{Virus} chain. If the \emph{Antibody} chain reaches the same length than the \emph{Virus} chain before the \emph{Virus} reaches the critical length $n$, then \emph{I} is cured, returning to their original state \emph{S} (or \emph{E}, respectively). Again, as mentioned before, the chain of replicating antibodies are not meant to represent the actual immune process of an infected host. 
Is important to note that \emph{T} units, as well as \emph{E}, are added to the system as perturbations, \emph{i.e.} they are introduced arbitrarily at a fixed time after the simulation has started.

The whole model results in a simulation system that contains 418 rules, composed by 6 different types of agents that, by recombination of their states of its sites, may form 1096 different species. These species are representing each possible element of the model, and in turn they can form a number of different complexes that cannot be accurately determined, exceeding $10^8$. An example of such complexes would be the following pattern:
\begin{verbatim}
Person(c~i,v!1,i,p~p3,l~c5),Virus(prv!1,nxt)
\end{verbatim}
which stands for \emph{a Person in the Infected state, bound to a Virus agent, not interacting with any other Kappa agent, who has a level 3 of panic and is located in city 5}.

%% file: parameters.tex
In addition to rules, the other important part of the model are the stochastic rates that allow the theoretical model to fit to available data. One of the essential parameters in our model is the encounter rate between two persons in the basal level of panic, $p_1$. This parameter governs the formation of complexes, so it influences many of our system's interactions. Considering cities with an average population density of 8700 persons per squared kilometer (see Simulations), we defined an average baseline of 10 encounters per day per person. This value is then used to define other encounter rates by multiplying it by different factors, accounting for the distribution rate of different interactions. Another important parameter is the rate of the resolution of each encounter, that has two parts. The first part is the rate that accounts for the duration of the encounter which, in most cases, follows an exponential decay law. The second part gives the probability to obtain each possible outcome. By multiplying these two parts, the total rate of encounters shows the expected average duration of 10 minutes for \emph{S-Z} encounters, and 5 minutes for \emph{E-Z} encounters. The output of these encounters exhibit ratios distributed proportionally to their respective probabilities.

The last important parameter is the panic constant, defined equal to $exp(1)$, that affects the reaction rates involving persons with propensity to increase their panic. To reflect that the panic level of persons involved in a reaction affects the rate of that reaction, we multiply the rate of this reaction by the panic constant to the power of the panic level of each \emph{Person} minus one. This is also applied to define the rate of encounters between a \emph{Person} and a \emph{Carrier}. On the other hand, in case of the encounter of a \emph{Person} with \emph{T}, the situation is the inverse, \emph{i.e.} we multiply the rate of this reaction by the panic constant to the power of one minus the panic level of each \emph{Person}. The following \emph{Kappa}-like notation exemplifies the encounter of an \emph{S} with a \emph{Z} (first interaction), the encounter of an \emph{S} with an \emph{I} (second interaction) and the encounter of an \emph{S} with a \emph{Carrier} (third interaction).

\begin{verbatim}
Person(c~s, i, p~p2, l~c3), Person(c~z, i, l~c3) -> \
Person(c~s, i!1, p~p2, l~c3), Person(c~z, i!1, l~c3) @ 'encounter-rate'*'panic'^1

Person(c~s, i, p~p3, l~c3), Person(c~i, i, p~p2, l~c3) -> \
Person(c~s, i!1, p~p3, l~c3), Person(c~i, i!1, p~p2, l~c3) @ 'encounter-rate'*'panic'^3

Person(c~s, i, p~p1, l~c5), Carrier(i, k, o~c5, d~c6,) -> \
Person(c~s, i!1, p~p1, l~c5), Carrier(i!1, k!2, o~c5, d~c6), km(p!2, o~c5, d~c6) \
@ 'carrier-encounter-rate'*'panic'^0
\end{verbatim}

In general, the interaction rate $f$ of an encounter is a function that depends of the rate constant and the panic level, as follows:

\begin{equation}
f(r,p)= \left\{ \begin{array}{lr} 
			r \prod_{i=1}^n\exp(1-p_i) & \textrm{in encounter with}\emph{T}\\
			r \prod_{i=1}^n\exp(p_i-1) & \textrm{all other encounters}
		\end{array}
		\right.
\end{equation}

Where, $r$ is the rate to be modified and $p=(p_1, p_2, \ldots, p_n)$ is a vector with the panic level of each \emph{Person}, affected by panic, in the left hand side of the reaction.
 
As stated before, the panic level also affects the outcome of encounters. For panic levels 1, 2 and 3, a $70\%$, $80\%$ and $90\%$ of the encounters are won by \emph{Z}, respectively. On the other hand, \emph{S} elements in the level 0 of panic are defeated only in the $40\%$ of the encounters.
\emph{E} units are added as perturbation to the system to destroy the threat, so they generate more encounters with \emph{Z}. This is reflected in a reaction rate $54$ times higher than the encounter rate of non-especialists \emph{Person} agents with \emph{Z}. Moreover, the probability of an \emph{E} winning the encounter is equal to $90\%$.

The rate of replication of the virus, $\delta$, is obtained by solving equation \ref{eq-det} to an incubation period of 8 to 16 hours \cite{survival}, which gives an $n$ equal to $13$ and a $\delta$ of $23.4880$. On the other hand, the response to treatment is set to a replication rate 26 times higher than $\delta$. Furthermore, the transport process for \emph{Carriers} is particularly fitted to each pair origin/destiny by equation \ref{eq-det} as the expected travel duration between cities at speeds ranging from 50 to 150 Km per hour. 

As we represent a hypothetical situation very close to a catastrophe, the accidental deaths, affecting \emph{S} and \emph{I}, are defined according to an exponential decay law that is fitted to an average death rate of one percent per year. \emph{E} and $Ei$ units may also have accidental deaths but, since it is assumed that they are prepared to confront a chaotic situation, their rate of accidental death is around 9 times lower. Moreover, we assume that a \emph{Z} can die after a certain time by starvation, therefore, an exponential decay reaction is fitted so that $90\%$ of \emph{Z} would be dead after 28 weeks \cite{week28}.

%% file: prekappa.tex
Even though \emph{Kappa} allows the definition of an important number of universal rules, \emph{i.e.} unique rules that apply for every city within our system, our model establishes several rules that apply in specific situations. For instance, while the elongation of the \emph{virus} and \emph{antibody} are universal rules, the kinetic rate of bimolecular interactions such as the encounter between an $S$ and a $Z$ depend on the area of each city (Figure \ref{cities-top}).  So, the need for a written rule for each different reaction becomes apparent. This issue leads to a \emph{Kappa} file with more than 700 lines of code for models composed by only 10 cities and 11 pairs of connected cities.
To facilitate the task of writing and editing a complex model such as a hypothetical Zombie outbreak in \emph{Kappa}, we defined a \emph{Kappa}-based syntax to express groups of location-specific rules in a compact way.  We call this intermediate language \emph{Prekappa} and we wrote and actively maintain a python script that reads a \emph{Prekappa} file and \emph{expands} it to a formal \emph{Kappa} file.

Understanding \emph{Prekappa} is straightforward after the examination of a few examples. Primarily, one defines the locations or compartments that will be modeled within the system as follows:
\begin{verbatim}
%loc: c1 0.5
%loc: c2 0.6
...
%loc: c0 0.8
\end{verbatim}
where c1...c0 are the location names, which will be used as internal states for the \emph{l} (location) site, and the following vector of numbers can have an arbitraty meaning. In this case, the number after the location name correspond to the area of each location relative to location c5.

In the following instructions, the modeler writes only one line that will be properly expanded to
one line for each location in a list of locations. Lists of locations can be created by:
\begin{verbatim}
%locl: Zone1 c1 c2 c3
\end{verbatim}
being the name \emph{all} reserved to refer to the whole set of locations previously defined.
For example, to write the rules of an encounter with a $Z$ applicable to all the cities, we use:
\begin{verbatim}
%expand-rule: all Person(c~s, i, p~p1),Person(c~z, i) -> \
Person(c~s, i!1, p~p1),Person(c~z, i!1) @ 10 / %loc[0]
\end{verbatim}
 and it will be expanded to formal \emph{Kappa} syntax as:
\begin{verbatim}
Person(c~s, i, p~p1, l~c1),Person(c~z, i, l~c1) -> \
Person(c~s, i!1, p~p1, l~c1),Person(c~z, i!1, l~c1) @ 10 / 0.5

Person(c~s, i, p~p1, l~c2),Person(c~z, i, l~c2) -> \
Person(c~s, i!1, p~p1, l~c2),Person(c~z, i!1, l~c2) @ 10 / 0.6

...

Person(c~s, i, p~p1, l~c0),Person(c~z, i, l~c0) -> \
Person(c~s, i!1, p~p1, l~c0),Person(c~z, i!1, l~c0) @ 10 / 0.8
\end{verbatim}
As seen, the use of \emph{Prekappa} allows us to declare a set of rules, even with different rate constants, in a single expression. In addition to locations lists, location matrices can be defined as:
\begin{verbatim}
%locm:
  TM  c1 c2 c3 c4 c5 c6 c7 c8 c9 c0
  c1  00 06 00 00 00 00 00 00 00 00
  c2  06 00 15 00 25 00 00 00 00 00
  ...
  c0  00 00 00 00 00 00 00 15 06 00
\end{verbatim}
 where TM is the matrix label and the cell values may have any arbitrary meaning. Location lists and location matrices may be used to expand any \emph{Kappa} expression: signature definitions, rules, \emph{init} statements, and variable declarations, including support for perturbations. So, TM is used to declare the signatures of our \emph{Carrier} agents and their initial number as follows:
\begin{verbatim}
%expand-agent: TM Carrier(i,k)
%expand-init: TM %cell Carrier()
\end{verbatim}
which results in:
\begin{verbatim}
%agent: Carrier(i,k,o~c1~c2~c3~c4~c5~c6~c7~c8~c9~c0, \
                    d~c1~c2~c3~c4~c5~c6~c7~c8~c9~c0)
%init: 6 Carrier(o~c1,d~c2)
%init: 6 Carrier(o~c2,d~c1)
%init: 15 Carrier(o~c2,d~c3)
%init: 25 Carrier(o~c2,d~c5)
...
%init: 15 Carrier(o~c0,d~c8)
%init: 6 Carrier(o~c0,d~c9)
\end{verbatim}
It is important to note how the expansion of the \emph{init} statements skipped the cells having values equal to zero in TM. The whole matrices can be expanded, indistinctly of their cells' values, by adding the flag \texttt{--full-matrix} to the command line when using the expander script.
Finally, a shortcut for writing chains of agents was defined so that a chain of, say, 10 \emph{kilometer} agents could be abbreviated as follows:
\begin{verbatim}
km(n!2),km(p!2,n!3),...,km(p!10)
\end{verbatim}
and then expanded as:
\begin{verbatim}
km(n!2),km(p!2,n!3),km(p!3,n!4),km(p!4,n!5),km(p!5,n!6), \
km(p!6,n!7),km(p!7,n!8),km(p!8,n!9),km(p!9,n!10)
\end{verbatim}

The use of \emph{prekappa} has dramatically reduced the size of our working files, from almost 1000 lines of code to less than 300.
Our expander tool is available at \texttt{https://github.com/DLab/expander}. Both, the \emph{prekappa} file and the resulting expanded \emph{Kappa} file, can be reviewed in the Supplementary Material.

%% file: simulations.tex
Simulations were run using \emph{KaSim} version 1.08 \cite{kappaorg}. KaSim implements the Gillespie's Stochastic Simulation Algorithm (SSA)  to solve a possible trajectory of the system's stochastic master equation \cite{gillespie}. According to Gillespie's approach, systems to be simulated are defined as a mixture of agents moving randomly in a given space, generating collisions between them that can lead to a reaction (interactions in our case) defined as a \emph{Kappa} rule \cite{danos2007}. Simulations were performed on the Levque cluster of the National Laboratory for High Performance Computing (NLHPC), Center for Mathematical Modeling (CMM), Universidad de Chile.

The simulation process was separated in two main parts. The first part was performed to determine the behavior of the system without countermeasures, namely $E$ and $T$ (from now on, the \emph{Basal Scenario}).  To do so, 1000 repetitions of the basal scenario were performed. These simulations were initiated with a total population of 53490 agents, spread proportionality to the area of each city (Figure \ref{cities-top}), corresponding to a population of near to 6 millions in a metropolitan area of 690$Km^2$ (similar to Santiago de Chile). To trigger the disease outbreak, 4 $Z$ elements were included in city $c1$. On the average of these simulations, we determined the critical times in which simulations reached the threshold of $5\%$ of $I$, to send the $T$ units,  and the $5\%$ of $Z$, to send the $E$ units, relative to the population of each city.  

To conduct the second part of the simulation process, 1100 tries of 25 scenarios were simulated, defined by a combination of different $E$ and $T$ units to be sent to each city. Deployed $E$ units were equal to $0.5\%$, $1\%$, $2\%$, $4\%$ and $8\%$ according to each city population, while dispatched $T$ units were enough to cover the demand for treatment of the $30\%$, $40\%$, $50\%$, $60\%$ and $70\%$ of the population, respectively (Table \ref{success-tab}). Each scenario was simulated for 300 days with the same initial conditions than the basal scenario. From these results, the simulations that exhibited representative results were considered for further statistical analysis. 

Due to the intrinsic stochasticity of the trajectories, each scenario was analyzed using scripts in \emph{R} language \cite{erre}.

%% file: results.tex
From the set of 1000 simulations of the basal scenario, in 974 the infection went on until complete elimination of the population. These simulations were considered as the representative ones. In the other 26 simulations the threat was neutralized early, without reaching a pandemic scenario (data not shown). Since one of the main goals of this work is the study of the dynamics occurring in a set of cities suffering from a catastrophic event, those 26 simulations were not considered for further statistical analysis. The average behavior of the other 974 simulations was used to determine the critical times to send the countermeasures, as shown in Table \ref{days-tab}. Those resulting times were consistent with the topology of the cities displayed in Figure \ref{cities-top}, in which the threat spreads from the nearest to farthest city with respect to the outbreak city ($c_1$). Furthermore, for every city, the critical day to send \emph{T} units resulted lower or equal to the time to send \emph{E} units, which is consistent to the incubation period of the disease (see Parameters). 

Table \ref{success-tab} shows the probabilities to overcome the disease as the ratio of successful simulations for each scenario, \emph{i.e.} those in which there are not infectious elements, namely \emph{I,D,Z} and \emph{Ei}, at the end of the simulation.  Analyses performed on this data show that regardless the number of available \emph{T} units, the inclusion of $0.5\%$ of \emph{E} is not enough to overcome the threat. The same can be seen in the scenario with $30\%$ of \emph{T} and $1.0\%$ of \emph{E}. Meanwhile, from $40\%$ to $70\%$ of \emph{T} and $1.0\%$ of \emph{E}, some successful simulations can be found. The first combination from which the situation becomes favorable is $30\%$ of \emph{T} and $8.0\%$ of \emph{E}, because a higher than $50\%$ probability to overcome the disaster was obtained. However successful, population survival in those simulations resulted quite low, as can be seen in Table \ref{survival-tab}, shown as a percentage relative to the total initial population.

Table \ref{survival-tab} presents the average percentage of survival \emph{S} with respect to the initial population at the end of each successful simulation (Table \ref{success-tab}).  As seen, survivors are scarce compared to the large amount of invested resources to rescue the cities from the zombie threat.  Surviving \emph{S} vary from  $0.23\%$ to $6.30\%$, when $40\%$ \emph{T} and $1\%$ \emph{E}, and $70\%$ \emph{T} and $8\%$ \emph{E}, were included in the simulation, respectively.  The scenario with $30\%$ of \emph{T} and $8\%$ of \emph{E} was selected for further analysis. This scenario shows a survival expectation of $3.75\%$.

As seen in Figure \ref{sidzet}, inset of panel A, \emph{S} elements present a short time without apparent change until day 3, followed by a quick decay, that stabilizes near to day 13, reaching only $3.75\%$ of the initial population. On the other hand, the average trajectories of \emph{I} and \emph{Z} present several breaks that correlate with the addition of \emph{T} and \emph{E} countermeasures, as shown  in Panel A. In the case of the \emph{Z} trajectory, breaks are notorious at days 4, 5, 6, 7, 8 and 9, being the highest at day 9 when the last 2 cities were intervened, as shown in Table \ref{days-tab}. The trajectory of \emph{I}, followed by the trajectory of \emph{D} present smoother breaks at critical days from 3 to 8.  

Figure \ref{sidzet}, Panel B, shows the total panic level of the simulation. As seen, panic level $p_1$ start to decrease almost at the beginning of the simulation while, at the same time, panic level $p_2$ start to increase at day 1 of the simulation. Notoriously, panic level $p_3$ start to increase at the mid of day 1, correlating with the appearance of \emph{I} as seen in Panel A. Thus, the peak of panic level $p_3$ strongly correlates with the maximum number of $I$ (Panel A). The lag shown in the trajectory's decay of \emph{S} (inset Panel A) is correlated with the increase of the panic level $p_2$, during the first three days. Inset of Panel B shows the marked growth of the special panic level $p_0$, reaching about 80 agents at the same time when the number of \emph{S} (inset Panel A) and \emph{I} (Panel A) reaches their minimum levels. 

As seen in Figure \ref{sidzet}, Panel C, the number of free \emph{Carriers} is stable from the beginning of the simulation until the mid of day 1, correlating with the appearance of panic level $p_3$. In accordance to this correlation, the number of free \emph{Carriers} reaches its minimum at a time when panic level $p_3$ becomes relevant compared to panic levels $p_1$ and $p_2$.

Taking into account that the system's global behavior emerge from the superposition of the inner dynamic of each city, we analyzed the average trajectories of some agents present in city $c5$ (Figure \ref{sidzet}, Panels D to F). As seen in the inset of Panel D, a marked lag in the trajectory of \emph{S} occurred from the beginning of the simulation until day 4, at a time when a rapid decay in the number of \emph{S} reach its minimum at day 10. This decay also correlates with the increase of panic level $p_0$, as seen in the inset of Panel E. 
The rapid decay in the number of \emph{S}, also strongly correlates with the appearance of \emph{I} elements, as can be seen in Panel D. As before, the trajectory of \emph{I} is followed by the trajectory of \emph{D}. On the other hand, as seen in Table \ref{days-tab}, \emph{T} countermeasures were added at day 6, meanwhile at the same time, \emph{E} elements were included in the simulation. These set of perturbations around day 6 produces a clear break in the tendencies of \emph{Z} and \emph{I} (Panel D). The trajectory of \emph{Z} respond instantly to the arrival of \emph{E} while the trajectory of \emph{I} present a certain delay with respect to the arrival of \emph{T}, as seen in Figure \ref{sidzet}, Panel D. The dynamics of \emph{S}, \emph{I}, \emph{D} and \emph{Z} was mainly developed between days 4 to 9, showing afterwards a rapid decay of \emph{S} until its stabilization at a minimum level at day 9. 

When analyzing the panic behavior in city  c5, its dynamics resulted clearly circumscribed between the beginning of the simulation, until day 10. From this day and on, no further changes in the level of panic was detected, as can be seen in Figure \ref{sidzet}, Panel E. Panic level $p_1$ experienced no changes until reaching day 2, followed from that point by a marked increase in panic level $p_2$. The permanence of the majority of the population in panic level $p_2$ is short -only 2 days- because at day 4, panic level $p_3$ raises as the predominant panic level of the population. Notoriously, changes in the panic level of the population that occur before day 4, proceed in absence of \emph{I}, \emph{D} and \emph{Z}, as seen in Panel D. Around day 5, the appearance of panic level $p_0$ resulted closely correlated with the maximum of panic level $p_3$, as seen in the inset of Figure \ref{sidzet}, Panel E. The maximum number of \emph{Persons} with panic level $p_0$ reach to 14 elements at day 8. The countermeasures effect on the panic level is reflected by changes in the trajectory of panic level $p_2$ at day 6, corresponding to the same day were \emph{T} and \emph{E} elements were added to the simulation.

Figure \ref{sidzet}, Panel F, presents the trajectory of the number of free \emph{Carriers} for city c5 during the simulation. As shown, the number of free \emph{Carriers} remain stable, showing a basal rate of occupancy, from the beginning of the simulation until day 2. From day 2, a marked decrease in the number of free \emph{Carriers} occurs until day 5, reaching its minimum number in the same day than the maximum of panic level $p_3$ (Panel E) and the beginning of the rapid decay in the number of \emph{S}, as seen in the inset of Panel D. 

%% file: discussion-model.tex
\subsection*{On the general assumptions of our model}
In spite that a zombie outbreak is a very unlikely situation, we decided to choose this model because of its resemblance to the spreading of infectious diseases in human populations. Moreover, the nature of this model gave us the ability to freely choose simulation parameters. According to popular mythologies, a zombie outbreak should impose a situation where people behave very unconventionally, being guided mostly by their panic, heavily focused towards personal survival, obeying population patterns beyond themselves. In such an event, people's situational awareness should be a key element to modulate the internal population dynamics. 
Following this idea and to study how the internal dynamics of this catastrophic scenario can be affected by people's panic, the ODE model in Figure \ref{sidzet-edo} is extended beyond the instantaneous change of the states regulated by a reaction rate. To do so, we considered all the possible outcomes for every encounter and their appropriate reaction rates. Therefore, the infectious process has two parts: the formation of the encounter and its resolution. Subsequently, we needed to estimate two parameters to adjust the model: the duration of the encounters and the chance of every possible outcome. These two elements are related in our model in order to define as many reactions as the number of possible outcomes from two interacting agents, as depicted in Figure \ref{sidzet-edo}.
By defining each rate as a function of the duration of the encounters, we can directly use the probability of each outcome and therefore, the resultant elements will be produced in quantities proportional to these probabilities. 
In our simulations, the number of effective encounters among agents was estimated as 10 per person per day, emphasizing that the encounter must be equivalent to a direct contact, strong enough to allow the transmission of the disease. Moreover, the duration of the encounters was defined as 10 minutes, except for the encounters involving \emph{E} units that span half that time to reflect the assumption that they are more effective in dealing with zombies. On the other hand, for the resolution of an encounter we must define a probability of contagion of the disease which, in our model, corresponds to the probability of losing a confrontation with a zombie. This parameter was set at 70\% for \emph{S} and \emph{I}, while \emph{E} units have only a 10\% chance to lose. This aims to reflect the lower capacity that people would have to react to the situation due to lack of preparation and the absence of countermeasures, which in turn, triggers a catastrophic scenario due to higher chances of getting the infection, before neutralizing the immediate threat. 
Interactions between people follows a similar mechanic of encounter-resolution but, in these cases, the resolution is separated in two steps: changes in the interacting agents and their subsequent separation. This gives rise to different possible outcomes from an encounter such as the occurrence of more than one reaction in the same complex or the separation of the complex without any change. These kind of three-step reactions are used in our model mainly to describe the spread of rumors and the encounters of civilians with special forces (\emph{E} units), both related to consequent changes in the level of panic.

The effect of panic on the internal dynamics of our model follows, as mentioned before, the notion that during catastrophic situations, mass or herd behavior is more determinant than the capacity of organization or rational decision-making. Specific values determining how panic affects the rate of reactions were arbitrarily defined.  However, they allowed us to rise or diminish those rates independently for each person depending on his/her emotional state, in contrast to a general approach that could change those rates for the whole system, yielding a richer system that accounts for more specificity and heterogeneity of the subjects. This is a highly desirable characteristic for epidemic models in which the heterogeneity of the susceptible population is a key element to be considered \cite{Bansal2007}.

In order to account for the individual disease progression, we decided to model in \emph{Kappa} the viral replication as seen in Figure \ref{kappa-rule}. \textit{A priori}, this implementation could increase the number of reactions and thus the computational requirements in a futile way. However, it is important to consider that a key factor to trigger a pandemic is the asymptomatic phase that renders travel restrictions and quarantine measures useless or inapplicable. Thus we decided to include an incubation period during which any infected person can travel freely between the cities encompassing our model. Even though the resolution of Equation \ref{eq-det} has nor biological neither epidemiological meaning regarding the number of elements needed for the expression of the symptoms, it allowed us to use a more precise adjustment to empirical data. As mentioned before, the mechanism of treatment involves the replication of \emph{Antibody} agents and has no relation with a real immune response. However, the resulting infective dynamics reflects a diminished effectiveness of the treatment, as the disease is more advanced, allowing us to control the time required to heal and the average effectiveness of treatments.

Regarding transport among cities, the use of \emph{Carrier} agents instead of a simple diffusion model of persons, allowed us to limit the maximum rate of travelling to reflect the finite transport capacity of a highway. This limitation was set to define a system susceptible to collapse if too many people wanted to travel, as in the case of the spreading of panic in catastrophic situations.

%% file: discussion-kappa.tex
\subsection*{On the use of the \emph{Kappa} language}

By looking at the local model shown in Figure \ref{sidzet-edo}, it would be expectable to solve it by traditional methods. However, this representation is the simplest way to explain the context and basic considerations of our model, without dealing with the rich set of details that a rule-based model could express. Starting from the basic model of Figure \ref{sidzet-edo}, our model was enriched by incorporating the description of the disease progression for every individual together with an approximation to model the effect of panic on people, giving rise to a complex system that would need more than 5700 variables to define the possible states of persons and their interaction with others, even in the same compartment. Interestingly, the number of variables needed to solve our system increase linearly with the amount of cities involved. So, although it would be possible to numerically solve such a large system, our main goal by using \emph{Kappa} was to reduce the complexity of the implementation without losing expressivity and the inherent richness of our model.
In addition to use a mass-action based model, treating the basic elements instead of the observables at a large scale, we needed a mechanistic approach such as rule-based model. On the other hand, under a Gillespie's approach, one must assume that the entities modeled are evenly distributed in space, so each compartment of a simulation -a city in our model- represents a well mixed reactor. To assume that the population of a city is evenly distributed may seem over simplistic, however, many of the approaches to model populations, such as ODEs, widely used in epidemiology and other sciences such as ecology, make the same assumptions yielding useful results \cite{Arino2006,Colizza2007plos}. Altogether, our model gave us a detailed view of the system at an exceptional level of granularity: the individual. In addition to deal with such a level of granularity, \emph{Kappa} allowed us to treat the combinatoric complexity by using patterns as generalizations of rules, significantly reducing the number of rules needed to describe the system. By implementing our \emph{Kappa} expander, the complexity of expanding \emph{Kappa} rules to a pseudo-explicit space, resulted in a simple task that allowed us to easily explore different and larger scenarios.

%% file: discussion-results.tex
\subsection*{On the results of our model}
By using SSA, the result of a simulation of a \emph{Kappa} model is one possible trajectory of the \emph{Stochastic Master Equation} that describes the system, seen as a set of reactions \cite{gillespie}. This represents a possible time-course for the evolution of the system given a certain set of initial conditions. Due to the stochasticity inherent to the SSA, the expected behavior of elements must be estimated from the average of many simulations. The initial set of simulations consisting in 1000 repetitions of the basal scenario, achieved two goals. The first was to determine, on the basis of the intrinsic properties of the non-intervened model -\emph{i.e.} basal scenario-, the critical times to send the countermeasures to each city. Secondly, allowed us to determine the number of simulations in which the threat was overcome before reaching a pandemic state. This situation arise as a consequence of the stochastic nature of the SSA simulation. The random outcome of the encounter between \emph{S} and \emph{Z}, combined with a very small number of initial cases of the disease, gives a 1\% chance of obtaining this outcome. We obtained 26 of these results from 1000 simulations, which can be expected given the chances, rendering its deletion from the results as valid.

By analyzing Table \ref{days-tab} one can notice the speed of the spread of the disease through the set of cities, observing the presence of \emph{Z} in every city by day 9. It is interesting to note that even though the vector of the disease does not have the ability to move between cities, the fact that the infected can travel is enough to trigger a pandemic event.

The definition of the critical times rises from the need to reduce the complexity of determining the best set of countermeasures that allow to overcome the catastrophic episode. Clearly, the variables to set are the optimal quantities of \emph{T} and \emph{E} and the times to send them, with the ability to set more than one time independently for each city. However, in the present work, we opted to fix the times and setting only one per city, to only focus on the effect that the quantities have over the outcome of the tries, as a function of the percentage of the of the initial population. The results in Table \ref{success-tab} can be interpreted as the probability of success of each scenario. It is clear that the value of the probability could be refined by considering a larger number of repetitions, but we decided to establish an average result that, first, was consistent with the aforementioned assumptions and secondly, with an adequate precision, while trying to keep computation times at reasonable level. The second of these requirements is the easier to support, given that by analyzing Figure \ref{sidzet} it is possible to note that the standard errors follow closely the average of the trajectories during all the simulation timespan. 

The consistence of our results is supported by the different behaviors seen in Figure \ref{sidzet}. First, the disaster triggered by the addition of a small number of zombies to the system has a large magnitude, decimating the population to a 3.75\% of the initial value (Table \ref{survival-tab}). A disaster of such magnitude, as expected, affects people physically and emotionally, generating chaos. This can be seen in the trajectories of the levels of panic, that presents a clear succession of the levels. This pattern shows a clear relation with the course of the pandemic, as seen by comparing panel A and B of Figure \ref{sidzet}; day 5, has most of the people at panic level 3, and it is the time at which the slope of the decay of \emph{susceptibles} becomes steeper. The mental state of the population also affects the occupancy of \emph{Carrier} agents, as seen in panel C of figure \ref{sidzet}. The amount of free \emph{Carriers} remains steady for a short time around 80 units, and starts to decay to a minimum of 30 as the number of people in higher levels of panic rises. This means that, as panic rises, more people is trying to travel. Finally, the amount of carriers rises again, possibly because there are fewer survivors to use them.

%% file: conclusion-perspective.tex
\subsection*{Conclusions}
By using as toy model a hypothetical zombie outbreak, this work was mainly focused in getting insights on the role that emotional factors may play on the population's behavioral adaptation, during the spread of a pandemic disease. To do so, we have implemented a rule-based model where the behavior of agents depend on their internal state of panic, measured as a function of the situational awareness. To date, many models studying the spread of infectious diseases assume that the population will not change its behavior in response to a disease outbreak \cite{Kiss:2010oq, Funk2010}. However, understanding the influence of populations' behavior on the spread of infectious diseases is key to plan, apply and improve any control measure\cite{Funk2010,Helbing:2000ve}. Systematic studies have demonstrated that public health measures such as school closures and quarantines actually produce behavioral adaptation\cite{Bootsma:2007fk,Hatchett:2007uq}. Despite centralized measures may force changes on the population's behavior, is important to consider that people may also experience self-induced changes. During the spread of infectious diseases it is common to observe actions aimed to reduce the risk of infection or to, at least, increase people's sense of security\cite{Bagnoli:2007ij}. Changes in the behavior of people to prevent sexually transmitted infections confirm that this is the case. An interesting example of this behavioral adaptation can be traced to the beginning of the 90's when the Gay population changed dramatically their sexual behavior because of the discovery of HIV \cite{Ferguson:2007ys}. Interestingly, risky behaviors in some communities have arised, as a result of recent studies demonstrating that a combined antiviral therapy for HIV can be used as a profilactic measure \cite{Krakower:2011zr}. During the SARS outbreak in 2003, people in Singapur and Hong Kong dramatically reduced travelling, weared mask in public and avoided contact with other people. According to this evidence, modeling approaches that doesn't consider the effect of emotional factors during the spread of an infectious disease, seems to be unable to capture the whole dynamics of the population. Thus, the comprehension of the underlying dynamics of the spread of an infectious disease in human populations, requires the understanding of the behavioral adaptation of both, the individual and the population being affected \cite{Ferguson:2007ys,Funk:2009dq,Funk2010}. As noted, during the spread of a contagious disease, people's behavior is modified producing changes in their contact network. In consequence, it is expected that basic models implementing static contact networks will produce inaccurate predictions\cite{Funk:2009dq,Funk:2010cr}. In order to overcome this limitation, our rule-base model incorporates a dynamical contact network that is generated throughout the simulation by the application of the Guillespie's SSA. The stochastic nature of this algorithm implies that this network can not be predicted from the initial parameters of the simulation. Moreover, the contact network of every agent is generated by its effective interactions through time.

In trying to address the effect of emotional factors, we were not focused on defining panic, rather, our motivation was to model how situational awareness, or information, may flow during the spread of the disease. This approach follows the idea that panic may be defined as a particular form of collective behaviour occurring in catastrophic situations \cite{Helbing:2000ve}. Interestingly, information diffusion can reduce the prevalence of infection if individuals avoid infection or seek treatment earlier\cite{Epstein:2008nx}. In order to incorporate these evidence in our simulations, \emph{S} agents in panic level \emph{p$_2$} or higher may use treatment without being infected, diminishing the number of available treatment to \emph{I} and \emph{E$_i$} agents. Again, by doing so, we have included in our simulations a behavior that depends on the individual situational awareness, but producing global effects on the population.
Despite some models assuming local clusters of information\cite{Salathe:2008qf}, the vast majority assume that the information on which people change their behavior, is available to everyone \cite{Funk:2009dq,Kiss:2010oq,Buonomo:2008tg}. On the contrary, in our model, information is taken from a mixture of the social and spatial neighborhood in which agents are immersed (local information), while information from other locations is obtained due to agents' traveling (global information).  Although classical \emph{Kappa}-based models assume that the system's dynamics occur in a well mixed reactor, we have extended this interpretation to model a reactor composed by different reaction chambers, or cities in our simulations. In addition, we have included information regarding the connectivity and the transportation system among cities, as has been successfully applied to other models to study disease spreading\cite{Colizza2006,VandenBroeck2011}. 
Unlike other models that study the influence of individual-based information\cite{Coelho:2009hc}, information availability to agents in our model, depends on the history of contact among agents during the simulation. Thus, behavioral adaptation for every agent will depend on the flow of both local and global information in a very unstructured manner. Interestingly, in socially or spatially structured models, information can occur in clusters which in turn can have strong effects on disease dynamics\cite{Salathe:2008qf}. In september 1994, an outbreak of pneumonic plague was reported in the city of Surat in western India \cite{Campbell:1995ly}. As a result of the highly unstructured information coming from unaware people, enhanced by the role of the media coverage\cite{Young:2008bs}, a panic explosion occurred in the Surat population. During a weekend, hundreds of thousands of persons fled away producing chaos and a huge number of casualties. Although later on, some cases were serologically confirmed as plague, the amount of human casualties largely surpassed those who died because of the disease. 

In addition to address the role of information flow, we have evaluated the effect of typical countermeasures of the zombie pop culture, considering the influence of a pseudo-explicit space where the population dynamics occurs. Considering that a zombie infection represents the spreading of a disease with a 100\% effective transmission, a short incubation period, no cure or remission, together with a profound emotional outcome for people being affected, the disease control is a highly complex task. Our simulations show that, under these circumstances, the population survival requires all, the availability and early dissemination of some treatment and the intervention of central authorities to control the disease spreading, by applying quarantines measures, and by eliminating the disease vectors. As expected from our implementation, the global dynamics of the infection progression on the population resulted primarily governed by the mechanistic description of local interactions. Notoriously, people's situational awareness resulted essential to modulate the inner dynamics of the system. These findings support the notion that, at least for pandemics with similar characteristics to this zombie outbreak, information moves faster than contagion. As demonstrated by pandemic episodes like SARS and, more recently, H1N1 (avian flu), governments and authorities should make extensive efforts to improve the situational awareness of the population being affected. These efforts should not imply information restriction. On the contrary, an early aware population will be capable to react both more promptly and properly, to any catastrophic situation.

%% file: figures.tex
\begin{figure}[!ht]
\begin{center}
\includegraphics[width=120mm]{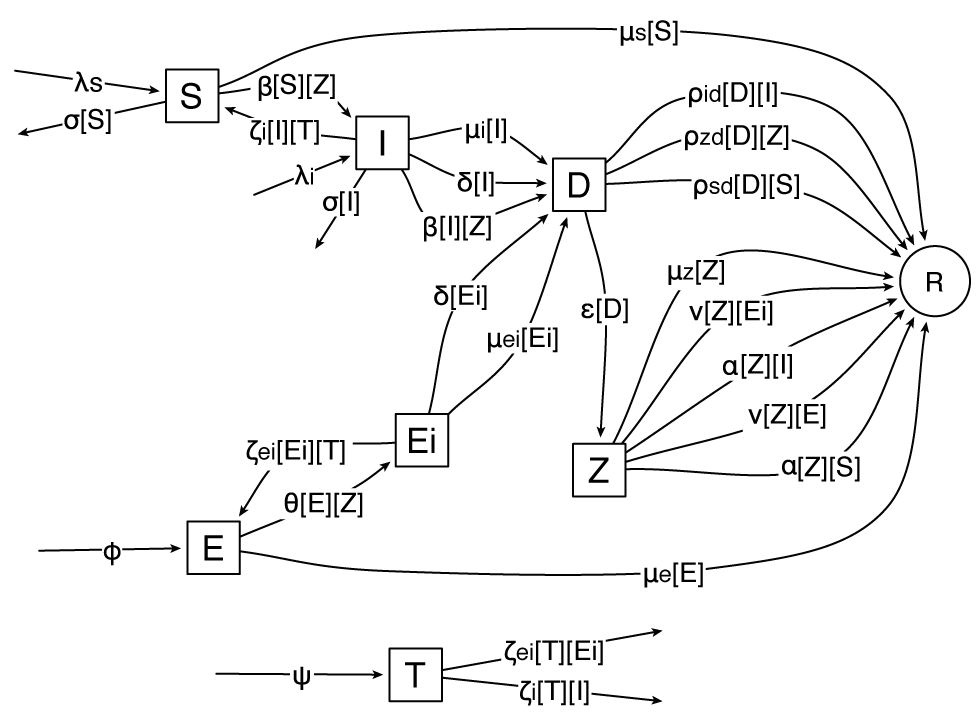}
\end{center}
\caption{
{\bf ODE model.}
Graphical representation of the local model for a Zombie outbreak using ordinary differential equations (ODE). Nodes represent different elements (defined as agents in our simulation) that describe the main states of our model and the arrows connecting them represents their possible transitions between states. The respective rate for every transition appears denoted as constants using greek letters, accompanied with the amount of every necessary element to produce the changes of states.
}
\label{sidzet-edo}
\end{figure}

\begin{figure}[!ht]
\begin{center}
\includegraphics[width=120mm]{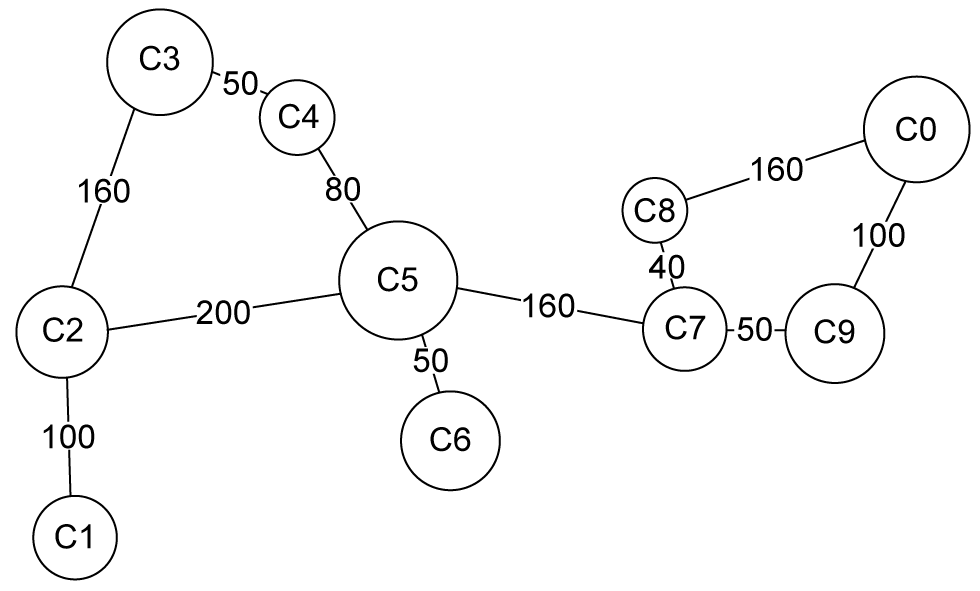}
\end{center}
\caption{
{\bf Topology of conectivity between cities.}
Our model considered a peuso-explicit space within the dynamics occur. The size of each circle is proportional to the area of the corresponding city and the length of the lines is proportional to the length of the roads connecting them, length that is indicated by the number on the lines.
}
\label{cities-top}
\end{figure}

\begin{figure}[!ht]
\begin{center}
\includegraphics[width=120mm]{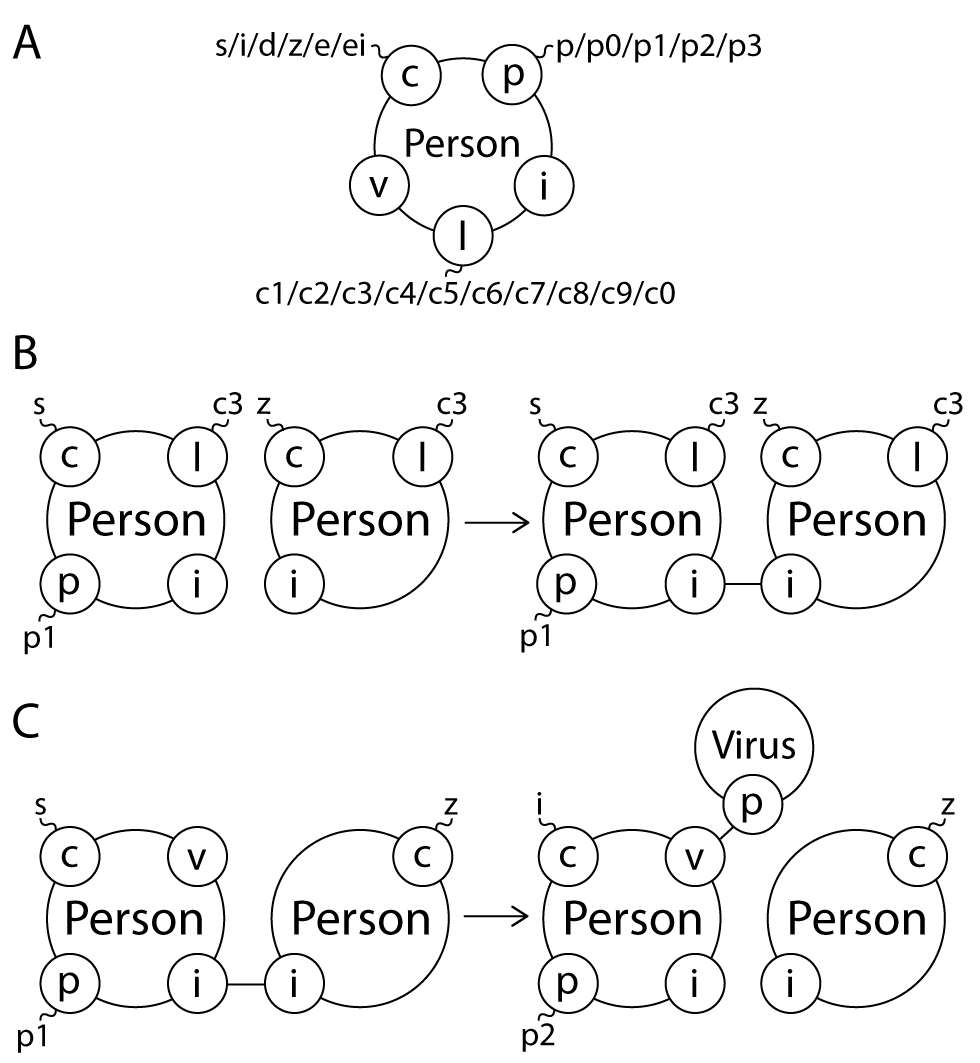}
\end{center}
\caption{
{\bf Graphic representation of \emph{Kappa} language.}
Panel A); Representation of the initialization of an agent \emph{Person} in \emph{Kappa}. All the five sites for any \emph{Person} are declared and every possible state for those sites is also shown. Panel B); Example of a \emph{Kappa} rule in which an \emph{S} and a \emph{Z}, represented by two persons with their site \emph{c} in states \emph{s} and \emph{z}, respectively, are located in the same city "c3" as denoted by the state \emph{c3} of its site \emph{l}. By interacting, these agents form an \emph{S-Z} complex by binding their sites \emph{i}. Panel C); A possible resolution of the \emph{S-Z} encounter in which the \emph{S} is infected by \emph{Z}, changing the state of his site \emph{c} from \emph{s} to \emph{i}, and increasing the level of panic by changing the state of \emph{p} from \emph{p1} to \emph{p2}. To complete the infection process, a \emph{Virus} agent is bound to \emph{I} and \emph{Z} is released. 
}
\label{kappa-rule}
\end{figure}

\begin{figure}[!ht]
\begin{center}
\includegraphics[width=130mm]{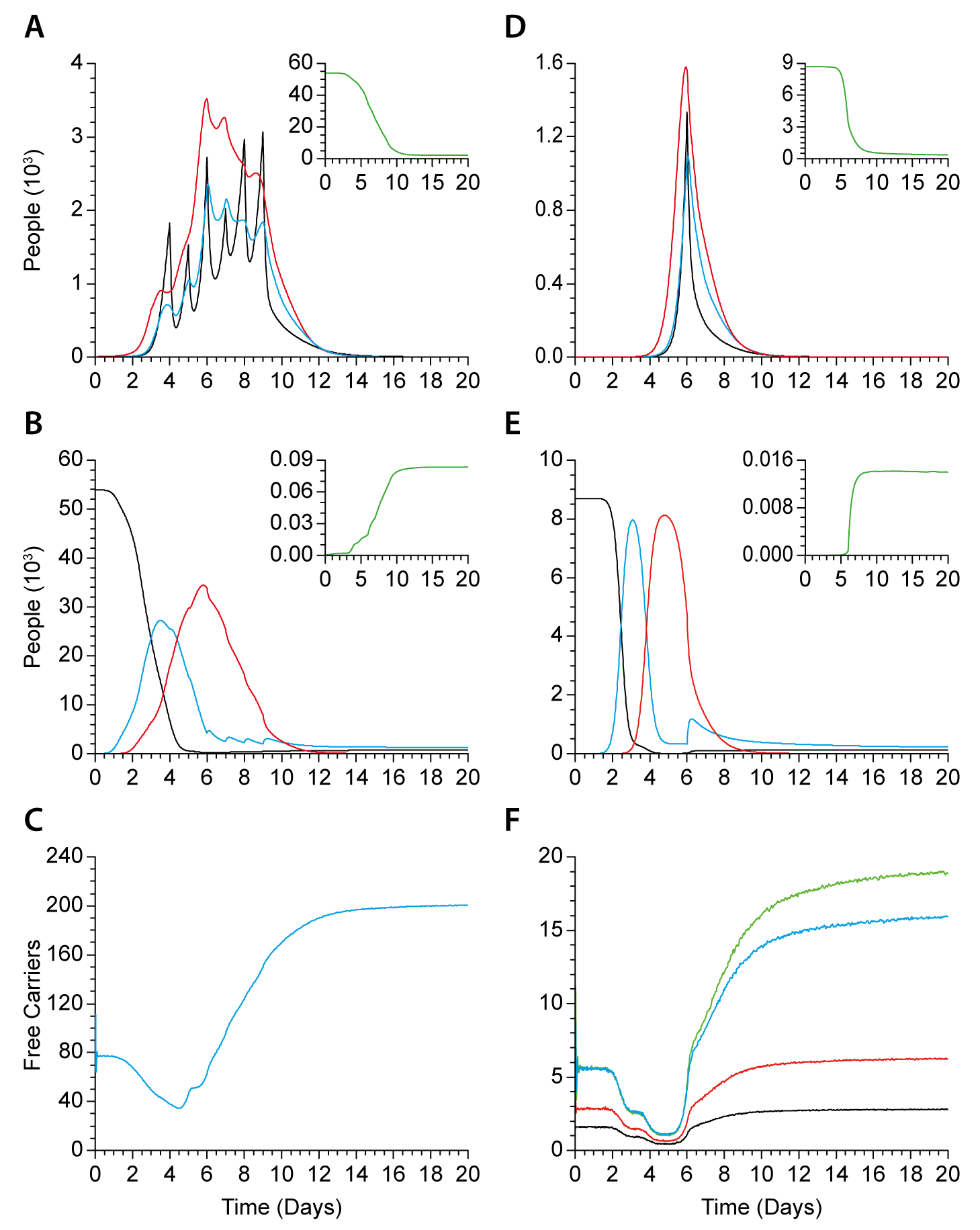}
\end{center}
\caption{
{\bf Average trajectories of the main elements.}
Statistical result of the 874 successful tries for the scenario with 30\% of \emph{T} and 8\% of \emph{E}. Left column represents global situation, meanwhile the right one describe the dynamic inside city c5. (A) Elements \emph{I} (red), \emph{D} (cyan) and \emph{Z} (black) are presented in the main graph, while \emph{S} (green) is presented in the inset. Both \emph{I} and \emph{Z} show breaks at times coincident with the addition of \emph{T} and \emph{E} to the system. (B) Levels of panic $p_1$, $p_2$ and $p_3$ (black, cyan and red, respectively) are presented in the main graph. The special level $p_0$ is presented in the inset (green). (C) Number of free Carriers showing the collapse of the transportation system between cities. (D) Elements \emph{I} (red), \emph{D} (cyan) and \emph{Z} (black) are presented in the main graph, while \emph{S} (green) is presented in the inset. (E) Levels of panic $p_1$, $p_2$ and $p_3$ (black, cyan and red, respectively) are presented on the main graph. The special level $p_0$ is presented in the inset (green). (F) Number of free Carriers at the connection paths from city c5 to cities c2, c4, c6 and c7 (green, red, black and cyan, respectively) present two major convexities that are consequent to the changes of panic levels.
}
\label{sidzet}
\end{figure}

%% file: tables.tex
\begin{table}[!ht]
\caption{
\bf{Predicted critical days}}
\begin{tabular}{ccccccccccc}
	\hline
	City & c1 & c2 & c3 & c4 & c5 & c6 & c7 & c8 & c9 & c0\\
	\hline
	Treatment & 3 & 4 & 5 & 6 & 6 & 7 & 7 & 8 & 8 & 9\\
	Exterminators & 4 & 5 & 6 & 7 & 6 & 8 & 7 & 8 & 9 & 9\\
	\hline
\end{tabular}
\begin{flushleft}
	Estimated days in which countermeasures were sent to each city. In the case of \emph{T}, the critical time to send the countermeasure was defined as the following day after the number of \emph{I} reached 5\% of the initial population for each city. \emph{E} units were sent to each city at the following day when the number of \emph{Z} reached 5\% of the initial population for each city. 

\end{flushleft}
\label{days-tab}
\end{table}

\begin{table}[!ht]
\caption{
\bf{Percentage of successful tries}}
\begin{tabular}{cccccc}
	\hline
	&\multicolumn{5}{c}{Treatment} \\
	\cline{2-6}
	Exterminators& 30 & 40 & 50 & 60 & 70\\
	\hline
	0.5 & 0.00 & 0.00 & 0.00 & 0.00 & 0.00\\
	1.0 & 0.00 & 0.09 & 0.19 & 1.13 & 1.39\\
	2.0 & 1.96 & 2.62 & 3.53 & 5.87 & 15.97\\
	4.0 & 11.06 & 13.18 & 16.00 & 16.99 & 20.54\\
	8.0 & 81.68 & 81.25 & 79.50 & 75.49 & 72.77\\
	\hline
\end{tabular}
\begin{flushleft}
	Percentage of successful tries for each scenario, in which there are not $I$, $D$, $Z$ or $Ei$ elements. All numbers in the treatment row and in the exterminator column represent their relative percentage to the initial population of each city. 
\end{flushleft}
\label{success-tab}
\end{table}

\begin{table}[!ht]
\caption{
\bf{Average percentage of survivals}}
\begin{tabular}{cccccc}
	\hline
	&\multicolumn{5}{c}{Treatment} \\
	\cline{2-6}
	Exterminators& 30 & 40 & 50 & 60 & 70\\
	\hline
	0.5 & 0.00 & 0.00 & 0.00 & 0.00 & 0.00\\
	1.0 & 0.00 & 0.23 & 0.51 & 0.67 & 0.78\\
	2.0 & 0.53 & 0.78 & 0.94 & 1.18 & 1.66\\
	4.0 & 1.48 & 1.86 & 2.27 & 2.76 & 3.01\\
	8.0 & 3.75 & 4.45 & 5.03 & 5.48 & 6.30\\
	\hline
\end{tabular}
\begin{flushleft}
	Average percentage of survival people for each scenario. Only the number of persons in successful tries is shown.
\end{flushleft}
\label{survival-tab}
\end{table}